\UseRawInputEncoding
\documentclass[%
 aip,
 amsmath,amssymb,
 reprint,%
]{revtex4-1}
\usepackage{graphicx}
\usepackage{color}
\usepackage{floatrow}
\usepackage{ragged2e}

\begin{document}

\title{Intrinsic synaptic plasticity of ferroelectric field effect transistors for online learning}%

\author{Arnob Saha}
 \altaffiliation[]{Authors contributed equally to this work.}
\author{A N M Nafiul Islam}%
  \altaffiliation[]{Authors contributed equally to this work.}
\affiliation{ 
School of Electrical Engineering \& Computer Science, The Pennsylvania State University, University Park, PA 16802, USA
}%

\author{Zijian Zhao}
\author{Shan Deng}
\author{Kai Ni}
\affiliation{ 
Microsystems Engineering Ph.D. program, Rochester Institute of Technology, Rochester, NY 14623, USA
}%


\author{Abhronil Sengupta}
\email{sengupta@psu.edu}
\affiliation{ 
School of Electrical Engineering \& Computer Science, The Pennsylvania State University, University Park, PA 16802, USA
}%

\begin{abstract}
{\small
Nanoelectronic devices emulating neuro-synaptic functionalities through their intrinsic physics at low operating energies is imperative toward the realization of brain-like neuromorphic computers. In this work, we leverage the non-linear voltage dependent partial polarization switching of a ferroelectric field effect transistor to mimic plasticity characteristics of biological synapses. We provide experimental measurements of the synaptic characteristics for a $28nm$ high-k metal gate technology based device and develop an experimentally calibrated device model for large-scale system performance prediction. Decoupled read-write paths, ultra-low programming energies and the possibility of arranging such devices in a cross-point architecture demonstrate the synaptic efficacy of the device. Our hardware-algorithm co-design analysis reveals that the intrinsic plasticity of the ferroelectric devices has potential to enable unsupervised local learning in edge devices with limited training data.}
\end{abstract}

\maketitle
\begin{figure}
\centering
\includegraphics[width=0.9\textwidth]{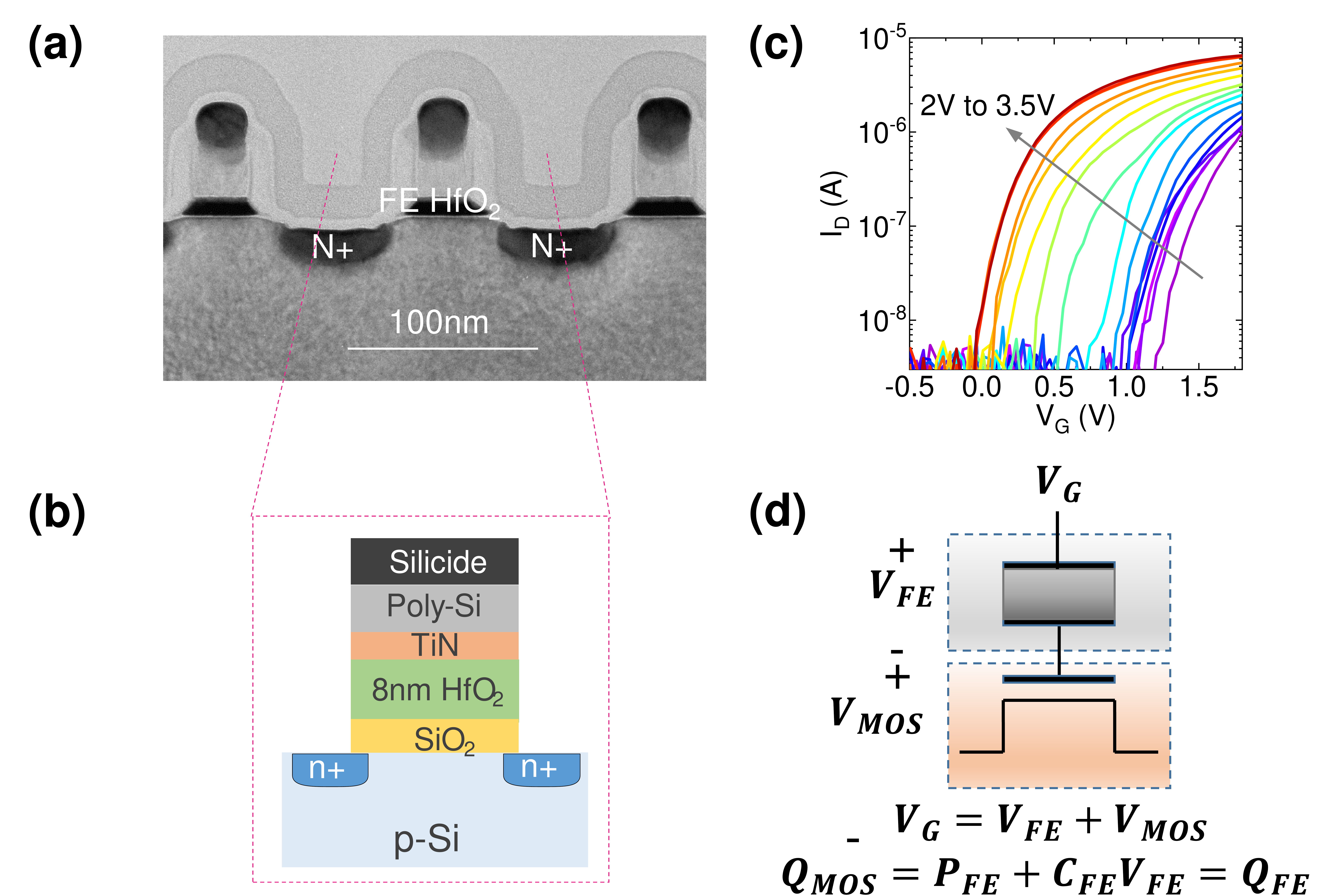}
\caption{\scriptsize{(a) TEM cross-section and (b) schematic structure of an industrial FeFET on 28nm technology node with $\textrm{HfO}_2$ as the ferroelectric layer. (c) Drain current ($I_D$) vs. gate voltage ($V_G$) characteristics of the FeFET. By applying programming pulses with increasing amplitudes, the threshold voltage $V_{TH}$, of the transistor can be gradually changed. For potentiation (depression), the FeFET threshold voltage decreases (increases). (d) FeFET device can be abstracted as a series connection of a ferroelectric capacitor on top of a MOSFET gate. Applied gate voltage ($V_G$) is divided across the ferroelectric layer ($V_{FE}$) and MOSFET ($V_{MOS}$). Due to charge continuity, charge in the MOSFET ($Q_{MOS}$) and in the ferroelectric layer ($Q_{FE}$), measured by the summation of ferroelectric polarization ($P_{FE}$) and the product of ferroelectric capacitance ($C_{FE}$) and voltage drop across the ferroelectric layer, are equal.}}
\label{fig1}
\vspace{-2mm}
\end{figure}

Harnessing the unique features of ferroelectric materials in essentially a device structure similar to that of a traditional transistor, Ferroelectric Field-Effect Transistor (FeFET) has the potential to expand the capabilities of current CMOS platforms to drive the innovations required in computing for emerging data-oriented applications. The recent discovery of ferroelectricity in CMOS compatible Hafnium Oxide \cite{boscke2011ferroelectricity,boscke2011phase} has made this a possibility due to its great scalability and superior energy efficiency. Previously, the incompatibility of perovskite based complex oxide ferroelectrics had put these devices by the wayside. The device, shown in Fig. \ref{fig1}(a-b), is characterized by a $\textrm{HfO}_2$ thin film ferroelectric layer replacing the conventional high-k dielectric in the gate stack of a MOSFET. Depending on its direction, the polarization either aids the formation of inversion or depletion/accumulation of the underlying semiconductor channel, hence modulating the threshold voltage of the FeFET. 

In addition, partial polarization switching can be induced by applying a voltage pulse train with either identical pulses or varying pulses of different amplitudes or pulse widths to the gate of the FeFET, such that a gradual tuning of the threshold voltage and subsequently the channel conductance can be realized (see Fig. \ref{fig1}(c)). This enables the application of FeFETs as multi-state non-volatile weight cells or synaptic memory elements – suitable for integration in dense memory cross-point arrays in neuromorphic systems. Cross-point arrays enable computation and memory update all within the same array itself, circumventing the von Neumann bottleneck of latency and energy overhead associated with moving data back and forth between logic and memory units. Additionally, the multi-level accumulative property of FeFET combined with their merged logic-memory functionality makes them very attractive for on-chip learning and designing adaptive intelligent systems for ubiquitous edge devices that are robust to the changing real-world environment while still remaining energy and area efficient. 

Previous works \cite{mulaosmanovic2017novel,jerry2017ferroelectric,sun2018exploiting} on exploring FeFETs as synaptic elements have primarily focused at the device level without considering system and algorithm level feedback. For instance, Refs. \cite{mulaosmanovic2017novel,jerry2017ferroelectric} abstract the FeFET synaptic device functionality driven by a pulse train where the pulsing scheme involves incremental pulse amplitudes. From an online learning perspective, such complex pulsing schemes would require us to read-out individual device states during the learning process and update each device individually by the appropriate pulse train, thereby requiring look-up table based approaches. The overhead of such schemes increase significantly with increasing network size, in addition to the already computationally expensive operations of online device programming events. Moreover, the complex learning schemes are also driven by the goal of achieving linearity in the programmable device states \cite{mulaosmanovic2017novel,jerry2017ferroelectric}. This work adopts the alternative route of leveraging the intrinsic plasticity characteristics of the FeFET device itself under single programming pulse excitation for learning algorithm formulation. This allows us to preserve the benefits of the core device to higher levels of circuit and system design abstraction. 

Designing brain-inspired adaptive intelligent systems that embrace the non-linear intrinsic plasticity of ferroelectric synapses instead of viewing it as a disadvantage will be critical to implement online learning with minimal peripheral overhead. In this work, we showcase the FeFET synaptic properties that can be exploited in a neuromorphic system including experimental characterization, device modelling and benchmarking, algorithm formulation based on intrinsic device physics along with the development of a device-circuit-algorithm co-simulation framework to assess the performance of the FeFET synaptic memory device at scale for a spiking neural network based adaptive learning system \cite{diehl2015unsupervised,sengupta2016hybrid,ambrogio2016unsupervised}.

The multiple state programming capability of ferroelectric materials occurs due to the presence of multi-domains, related with the lateral grain size \cite{lee2021domains}. The number of domains (and available programming states) therefore increases with increasing device size. Since the coercive voltage varies spatially across domains, the switching voltage of a particular domain varies in proportion with the corresponding coercive voltage. Grain size and grain orientation of the ferroelectric material are some of the major factors that influence the spatially inhomogeneous distribution of the coercive voltage \cite{mulaosmanovic2017switching}. Upon application of a programming gate voltage, discrete probabilistic switching events of a subset of the domains take place resulting in the threshold voltage of the FeFET changing gradually. 

\begin{figure*}[t]
\centering
\includegraphics[width=\textwidth]{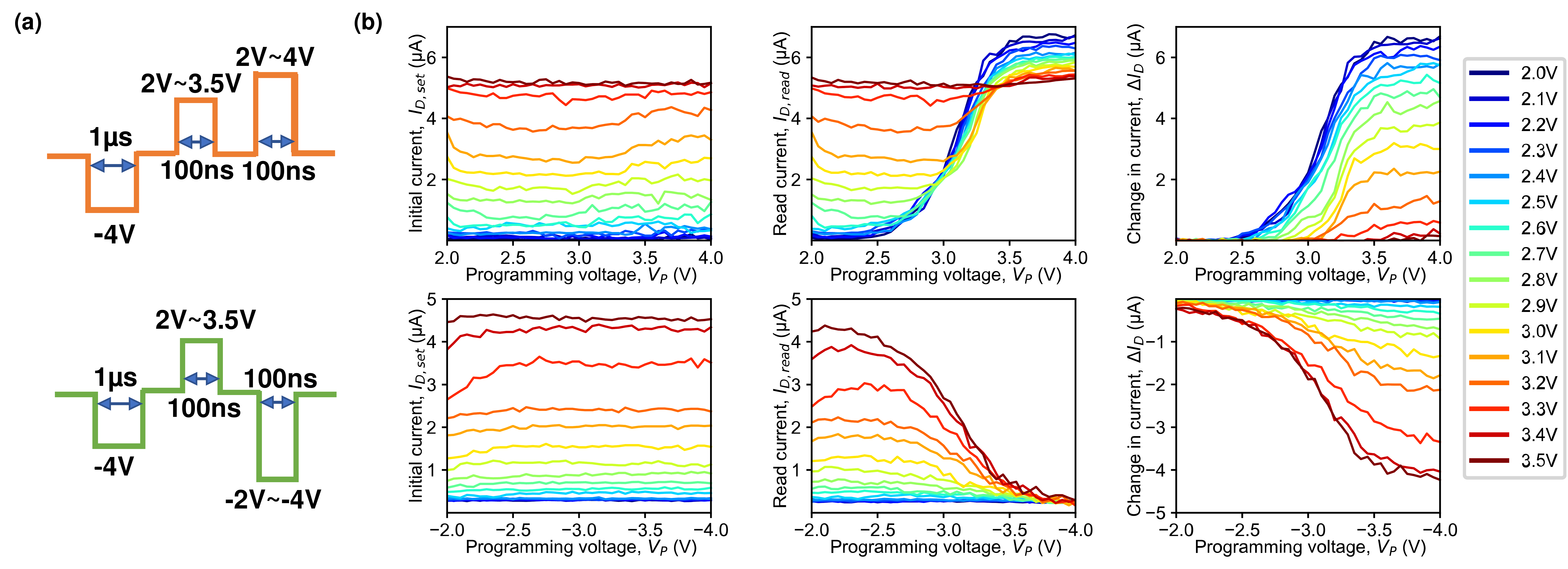}
\caption{\scriptsize{(a) Pulsing scheme for potentiation (orange) and depression (green) with a reset pulse of magnitude = $-4V$ and pulse duration = $1 \mu s$, set pulse of magnitude = $2V$ to $3.5V$ and pulse duration = $100ns$ and programming pulse of magnitude = $2V$ to $4V$ ($-2V$ to $-4V$) for potentiation (depression) and pulse duration = $100ns$. (b) Experimentally measured drain current values have been plotted with respect to the programming voltage amplitude ($V_P$). Top left (bottom left) plot shows the measured initial current ($I_{D,set}$) after the set pulse for potentiation (depression) for different set pulse voltages. The curves are almost horizontal as it is independent of the programming voltage. Top middle (bottom middle) plot shows measured read current ($I_{D,read}$) after the programming pulse during read condition for potentiation (depression). Top right (bottom right) plot shows the change in current ($\Delta I_D$) for potentiation (depression).}}
\label{exp}
\vspace{-2mm}

\end{figure*}

In order to understand the synaptic plasticity characteristics of the FeFET device, we measured the change in channel conductance of the device in response to programming voltages of varying amplitude. The synaptic plasticity/weight change is equivalent to the FeFET conductance change. For a given programming pulse amplitude, FeFETs with lower conductance undergo a greater increase in conductance after the ``write" operation since lower conductance is achieved by a larger polarization pointing at the gate metal, hence more are available to switch during the positive ``write" pulses, which therefore induces a greater conductance change. 
The reverse is true for the depression operation where the device conductance is decreased due to the application of a negative programming voltage. To capture this effect in our device modelling, we performed experimental measurements for a $1\mu m \times 1\mu m$ FeFET device triggered by a pulsing scheme, as shown in Fig. \ref{exp}(a) which consists of a reset pulse, a set pulse and a programming pulse. The FeFET is based on an industrial $28nm$ high-k metal gate (HKMG) technology \cite{trentzsch201628nm}. The device features an $8nm$ thick doped HfO$_2$ ferroelectric layer and $\sim 1nm$ SiO$_2$ native oxide. During characterization, a reset pulse of magnitude $-4V$ is applied to reset all the domains of the ferroelectric layer to negative polarization states. The device is initialized to different intermediate conductance states by changing the set pulse voltage from $2V$ to $3.5V$ with increments of $0.1V$. The device plasticity characteristics is then evaluated by the application of a subsequent programming pulse. Finally, measurements are performed for different programming voltage amplitudes ranging from $2V$ to $4V$ with $50mV$ steps. Drain current values after the set pulse ($I_{D,set}$) and the programming pulse ($I_{D,read}$) are measured for different set pulse voltages representing different intermediate states of the device. Changes in current ($\Delta I_D=I_{D,read}-I_{D,set}$) as a function of the programming voltage are calculated to reflect the synaptic conductance change for potentiation and depression (Fig. \ref{exp}(b)). 

\begin{figure}[t]
\centering
\includegraphics[width=\textwidth]{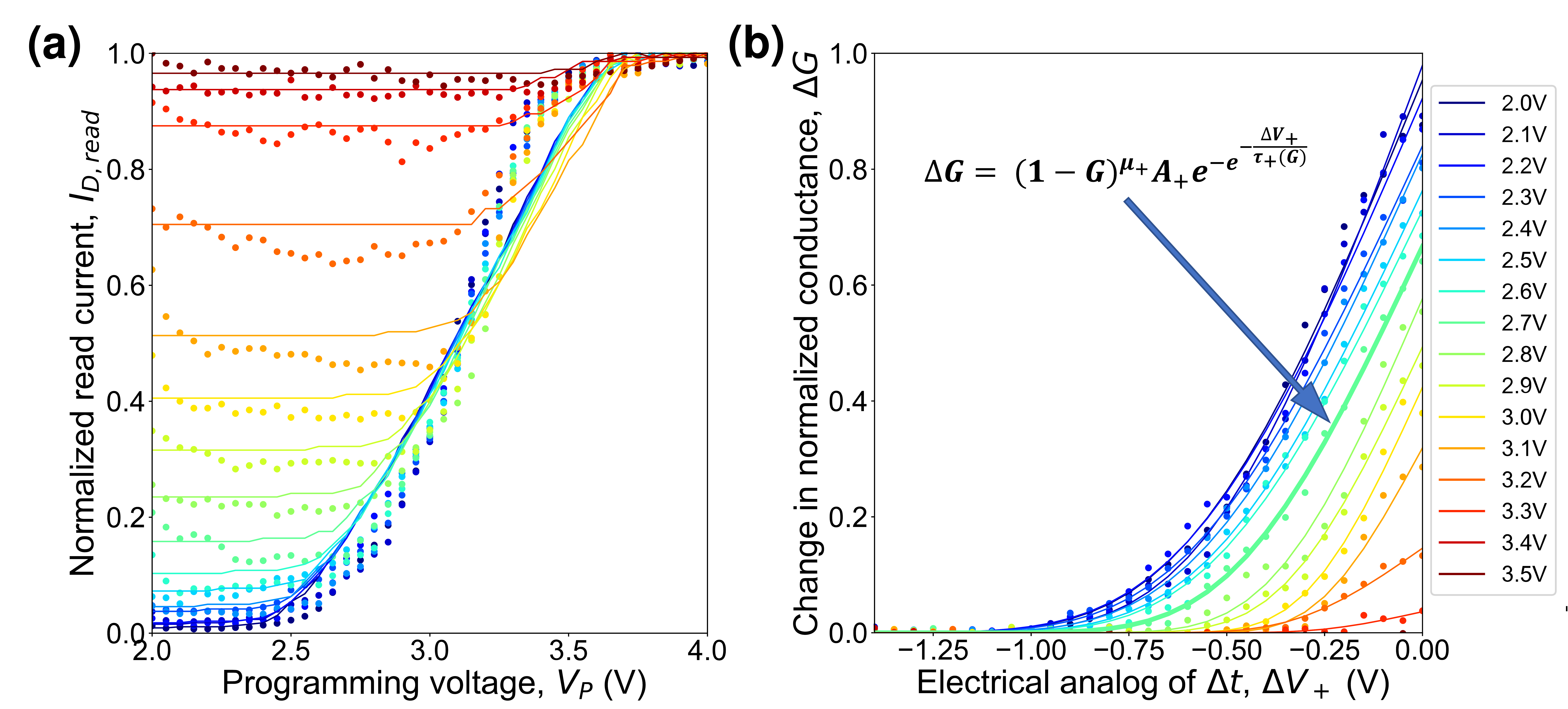}
\caption{\scriptsize{(a) Normalized read current ($I_{D,read}$) variation with the amplitude of the programming pulse have been plotted for different set pulse amplitudes. The calibrated model (solid line) matches well with the experimental data (scattered data points). (b) Compact modelling of the device STDP characteristics is achieved by abstracting the variation of the change of normalized FeFET conductance vs. the electrical analogue of spike timing delay, $\Delta V_+ = V_P - V_{o+}$, where $V_{o+} = 3.4V$.}}
\label{3ab}
\vspace{-2mm}
\end{figure}

To develop an experimentally calibrated device model for our device-algorithm co-simulation framework, we utilized a Monte Carlo algorithm \cite{alessandri2019monte,deng2020comprehensive} to calculate the switching probability of domains at each time step. The ferroelectric behavior modelling framework considers that the ferroelectric layer consists of multiple independent domains, polarized in either one of the two stable orientations. Under a constant electric field, $E_{fe}$, the nucleation time constant, $\tau$ is independent of time \cite{alessandri2019monte}. If a domain switches within a certain time step, $\Delta t$, the domain switching probability, $p_i$ can be expressed as \cite{alessandri2019monte},
\begin{equation}
\label{switching probability}
p_i (t_s < t+\Delta t\ |\ t_s > t) = 1-e^{({\frac{t}{\tau_i}})^\beta - ({\frac{t+\Delta t}{\tau_i}})^\beta}
\end{equation}
where, $\beta$ is the shape parameter of the probability distribution. $t_s$ is the switching time of the $i$-th domain considering that the domain has not switched before $t$, and $\tau_i$ is the switching time constant of the $i$-th domain. In case of a switching event, the state, $s_i(t)$, of the $i$-th domain gets altered. Total polarization, $P_{total}$, can be measured by taking the summation of the states of all the $N$ number of domains present in the ferroelectric layer,
\begin{equation}
\label{total polarization}
P_{total}(t) = \frac{P_r}{N} \sum_{i=1}^{N} s_i(t)
\end{equation}
where, $P_r$ is a constant polarization for each domain. In our case, since the applied electric field is temporally varying ($E_{fe}(t)$), the switching time constant becomes a function of activation field, $E_{a,i}$ and applied electric field, $E_{fe}(t)$. $\frac{t}{\tau_i}$ in Eqn. \ref{switching probability} can be replaced by the history parameter, $h_i(t)$ \cite{alessandri2019monte} of the form, 
\begin{equation}
\label{history parameter}
h_i(t) =\int_{t_0}^{t} \frac{dt'}{\tau_i(E_{fe}(t'),E_{a,i})}
\end{equation}
The history parameter, $h_i(t)$, increases till the domain flips. This growth captures the accumulative behavior of the ferroelectric layer and therefore the switching probability of a domain can be expressed as,
\begin{equation}
p_i(t_s < t+\Delta t\ |\ t_s > t) = 1-e^{(h_i(t))^\beta - (h_i(t+\Delta t))^\beta}
\end{equation}
By tracking the polarization behavior, the model is then self-consistently solved with transistor charge-voltage equations (see Fig. \ref{fig1}(d)) to obtain FeFET device characteristics. As shown in Fig. \ref{3ab}(a), the normalized read current for different device intermediate states matches well with the corresponding experimental data.

We studied the intrinsic adaptive synaptic characteristics of the FeFET in the context of unsupervised Spike Timing Dependent Plasticity (STDP) \cite{bi1998synaptic} in neuromorphic systems. In such a system, “spikes” are propagated through the network and the synaptic weights are modified by STDP according to the time difference between spiking events of the pre- and post-synaptic neurons. Similar to the way the biological brain works, networks employing such spikes and synaptic learning rules are referred to as Spiking Neural Networks (SNN) \cite{sengupta2019going} and are gaining greater interest owing to their bio-plausibility, energy-efficiency and hardware friendly mechanism \cite{aimone2021roadmap}. The unsupervised, local, few-shot clustering capability of STDP is an ideal fit for on-chip embedded intelligence in resource constrained edge devices with limited labelled data where it may not be always possible to transmit information to the cloud for real-time processing. In order to envisage the implementation of STDP in FeFET devices, we consider the devices to be programmed by an adaptive programming voltage which is tuned in accordance to the timing delay between pre- and post-synaptic neuron spikes. Interestingly, the $\Delta I_D$ plots in Fig. \ref{exp}(b) reveal that the characteristics saturate at a particular voltage, $V_{0+} (V_{0-})$ for potentiation (depression) for all the intermediate states, thereby indicating a saturation voltage beyond which the device is programmed to the other state. Let us discuss the STDP implementation for the potentiation phase. Similar discussions are also valid for depression. STDP corresponds to the change in synaptic weight ($\Delta w$) as a function of the time difference ($\Delta t$) between post-synaptic neuron and pre-synaptic neuron spikes. The weight change is maximum at $\Delta t = 0$ and decays with increasing $\Delta t$, thereby promoting temporal correlation. This saturation voltage ($V_{0+}$) therefore corresponds to the voltage that is applied to the FeFET corresponding to the $\Delta t=0$ condition. Assuming a direct linear correspondence between applied programming voltage and spiking time-delay (to be implemented by appropriate peripheral operation discussed later), we abstract the STDP synaptic weight change, $\Delta w$ (electrical analogue: normalized device conductance change, $\Delta G)$, as a function of the time-delay between spikes (electrical analogue: programming voltage measured with respect to the saturation voltage applied corresponding to $\Delta t = 0$, $\Delta V_+ = V_P - V_{0+}, \propto \Delta t$) as,
\begin{equation}
\Delta G=
\begin{cases}
\phantom{}(1-G)^ {\mu_{+}}\times A_+\times e^{-e^{\frac{-\Delta V_{+}}{\tau_+(G)}}}, & \text{$\Delta t >0$}\\
    (G)^ {\mu_{-}}\times A_-\times e^{-e^{\frac{-\Delta V_{-}}{\tau_-(G)}}}, & \text{$\Delta t <0$}
      
\end{cases}
\label{STDPeq}
\end{equation}
where, $A_+$ and $A_-$ are constants, $\mu_+$ and $\mu_-$ refers to the degree of dependence of $\Delta G$ on current conductance state $G$, $\tau_+$ and $\tau_-$ represent the STDP time constants. The double exponential nature of the phenomenological model was based on similar switching characteristic modelling of other memristive technologies \cite{wijesinghe2018all}. The FeFET STDP compact model parameters are extracted from our experimentally calibrated device simulation framework ($A_{+/-}=2.91/-2.52,\mu_{+/-}=2.5/1.5,\tau_{+/-}(G)=0.57-0.76G/-1.54-0.79G$). The conductance dependent STDP behavior (See Fig. \ref{3ab}(b)) is reminiscent of a special type of STDP learning rule, known as the multiplicative STDP, where the weight update is not only dependent on activity but also a function of the weight of the synapse, i.e., the degree of potentiation of stronger synapses is weaker compared to their degree of depression and vice versa for weak synapses \cite{rubin2001equilibrium,van2000stable}. Such a multiplicative plasticity mechanism, enabled by the intrinsic FeFET device physics, is critical for ensuring stability and self-adaptation characteristics of SNNs \cite{prezioso2016self,rubin2001equilibrium,van2000stable}.

\begin{figure}[t]
\centering
\includegraphics[width=\textwidth]{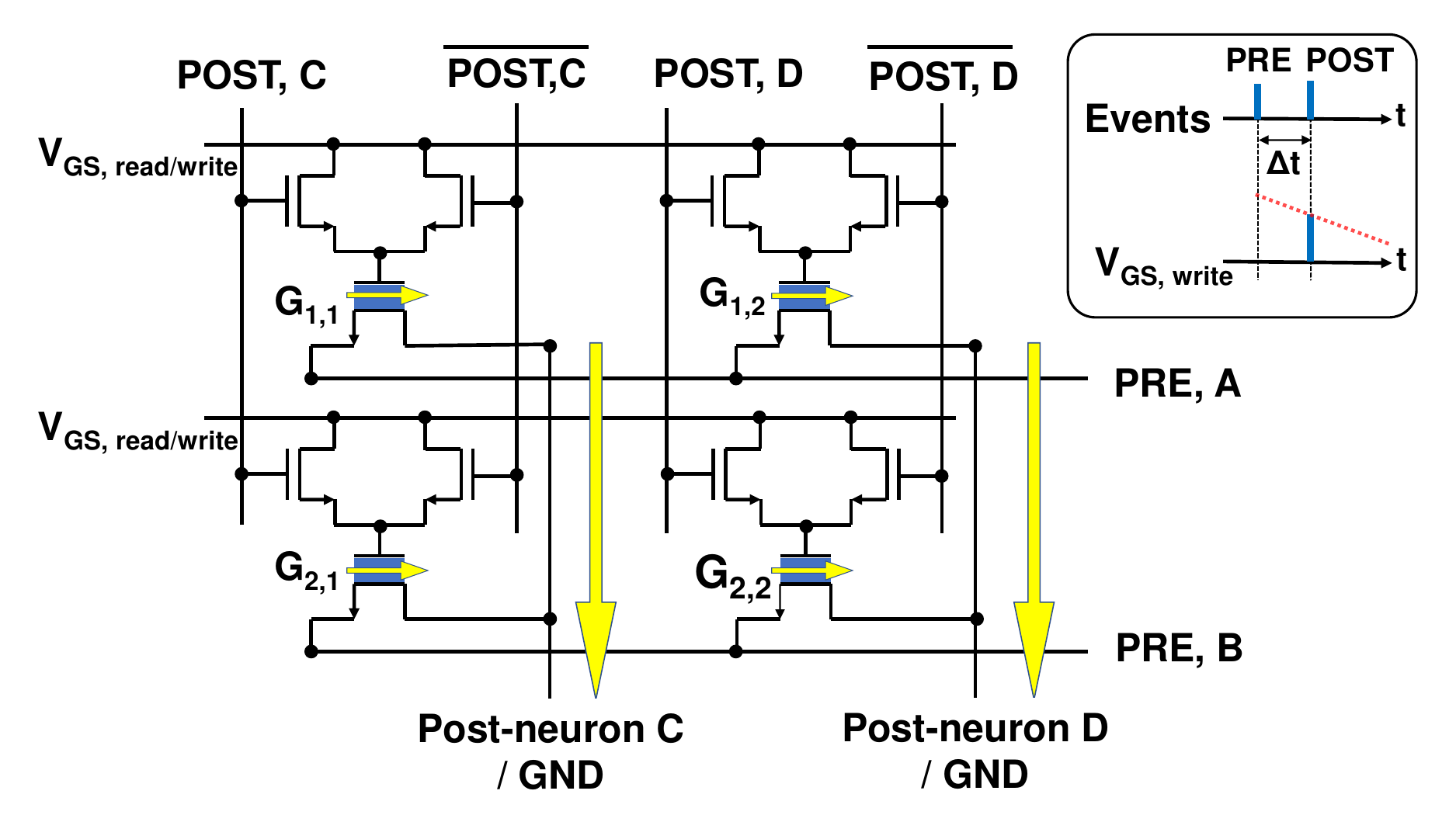}
\caption{\scriptsize{Cross-array architecture of FeFET devices. The incoming PRE spike signals are modified by the device conductance and are summed along the column following Kirchhoff's current law. The post-synaptic neurons integrate over the current and spike when the membrane potential reaches the threshold. The POST signals are then activated and during programming apply the appropriate amplitude of $V_{GS,write}$ (proportional to $\Delta t$, shown in inset) across the device to modulate the plasticity.}}
\label{cross}
\vspace{-2mm}
\end{figure}

\begin{figure*}
\centering
\includegraphics[width=1.0\textwidth]{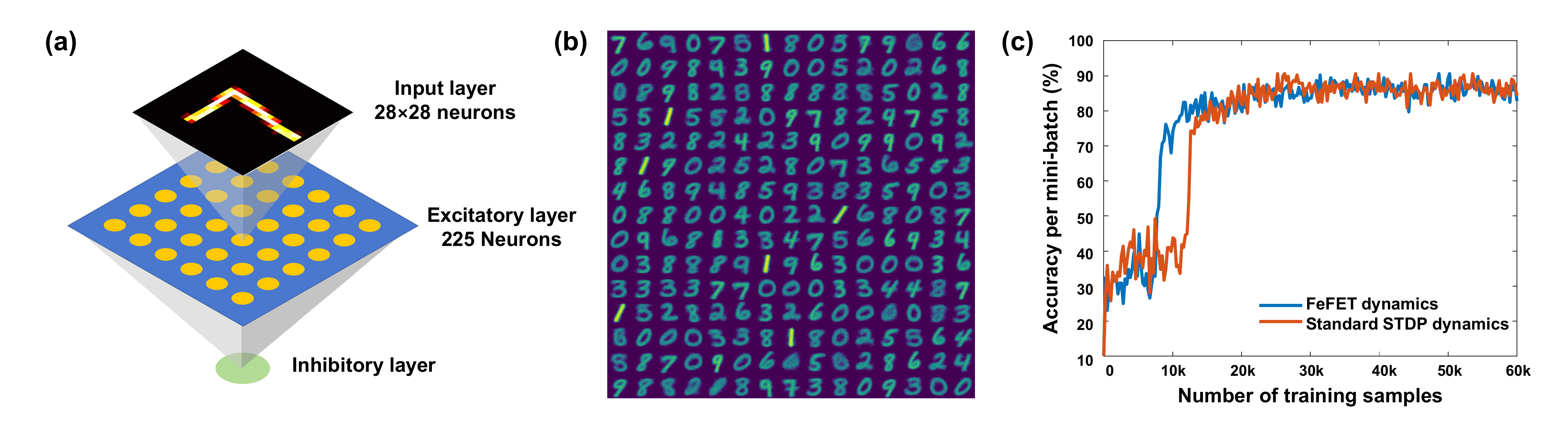}
\caption{\scriptsize{(a) Our SNN architecture consists of an excitatory neuron layer with lateral inhibitory connections for unsupervised learning. The synaptic weights joining the input and excitatory layer are implemented using FeFET devices and undego STDP learning. (b) FeFET synaptic array conductance states after training on the MNIST dataset for the 225-neuron network, (c) Accuracy per mini-batch during training with respect to the number of training samples. We observed a sharp rise in accuracy as the network activity decreased and weights became learnt. Further updates mostly fine-tuned the weights. It can be seen that this convergence in learning occurs with 33.3\% fewer training samples while using the FeFET dynamics compared to the standard STDP learning rule.}}
\label{network}
\vspace{-2mm}
\end{figure*}

Unlike prior complex programming voltage waveform engineering schemes for STDP implementation \cite{serrano2013stdp}, we adopt a simple cross-array operation such that the core device benefits can be preserved at circuit and system levels. The proposed cross-array structure is shown in Fig. \ref{cross} connecting PRE (transmitting)-synaptic neurons A, B to POST (receiving)-synaptic neurons C, D. The FeFET devices are connected with access transistors to decouple the ``write" and ``read" current paths. We would like to mention here that this decoupled ``write" and ``read" paths are critical for online learning to ensure independent optimization of the two operations. We program the FeFET devices every time the post-synaptic neuron fires ($t_2$) and activate the $V_{POST}$ control signal. Let us first discuss the potentiation, i.e. the positive STDP learning window ($\Delta t>0$). At the beginning of the positive time-window ($t_1$), the $V_{GS, write}$ is linearly decreased such that whenever the access transistor is on, the necessary programming voltage for potentiation (applied at the gate terminal of the FeFET) is tuned in accordance to $\Delta t = t_2 - t_1$. The negative learning window can be implemented by a simple extension of the above scheme \cite{sengupta2016hybrid,srinivasan2016magnetic} by using a negative $V_{GS,write}$ during the programming process.




We evaluate the performance of these synaptic devices and their multiplicative dynamics in a network learning scenario. An experimentally calibrated device-algorithm co-simulation framework was devised in BindsNET \cite{hazan2018bindsnet} (a PyTorch based package) to train an unsupervised SNN \cite{diehl2015unsupervised} on a standard handwritten digit recognition problem based on the MNIST dataset\cite{lecun1998mnist}. The network, shown in Fig. \ref{network}(a), consists of an excitatory layer of neurons that receives the weighted summation of spikes from the inputs. The analog pixel intensities of the image are mapped to the firing rate of the input spike trains through a Poisson process. The excitatory layer can be therefore mapped to a cross-array of FeFET synapses, as shown in Fig. \ref{cross}, where the devices are programmed in an online fashion based on spike timings. The PRE-synaptic neurons correspond to the input neurons while the POST-synaptic neurons correspond to the excitatory layer neurons. In order to promote competition, the excitatory neurons are characterized by lateral inhibition effects by preventing other neurons from spiking whenever a particular excitatory neuron fires. Additionally, to ensure no single neuron dominates the firing pattern, homeostasis is implemented through adaptive thresholding where the neuron threshold increases every time a neuron fires. Lateral inhibition and homeostasis effects can be implemented by simple peripheral circuitry \cite{sengupta2016hybrid} of the FeFET synaptic cross-array.

Our network simulations consisted of an excitatory layer of $225$ spiking neurons characterized by Leaky Integrate and Fire (LIF) dynamics with intrinsic FeFET synaptic plasticity (see Fig. \ref{network}(a-b)). In order to assess the impact of the intrinsic device characteristics on the learning process, we performed a comparative simulation with standard STDP rule. As shown in Fig. \ref{network}(c), the multiplicative STDP behavior of the devices resulted in significantly faster convergence of the network, which is advantageous from the perspective of few-shot learning and minimizing costly programming events in the FeFET cross-array. We postulate that this advantage is mainly due to the self-adaptive behavior of multiplicative STDP, as mentioned before. As the weights are randomly initialized at the beginning of the training process, synaptic weights which are further away from their eventual converged values experience a larger synaptic change (potentiation or depression) compared to weights that are relatively close to their final magnitudes. It is worth mentioning here that this observation is specific to the pattern recognition system studied in this work and may not be generalizable (for instance, in networks exhibiting transmission delays \cite{asl2017dendritic,asl2018delay}). Our FeFET network simulations yielded a test accuracy of $87.82\%$ on the MNIST test dataset  (Fig. \ref{network}(c)) and required $33.3\%$ fewer training images than a standard STDP learning framework. Our recognition accuracies are at par with idealized software accuracies observed for networks of similar size \cite{diehl2015unsupervised}. The maximum programming energy consumption of the device during the learning process was evaluated to be $29.6 fJ$, which is significantly lower than other competitive technologies like phase change memories, spintronics, resistive random access memories, among others \cite{jackson2013nanoscale,kuzum2011nanoelectronic,li2014activity,jo2010nanoscale,nishitani2013dynamic,ramakrishnan2011floating,vijay_6,jerry2017ultra,sengupta2017encoding,sengupta2016proposal}.

In order to assess the impact of limited programming resolution typically observed in such emerging technologies \cite{sun2018exploiting,luo2019benchmark}, we also performed network simulations where the synaptic weight update was discretized. We used $6$-bit resolution in our work in accordance with prior literature \cite{wang2020investigating}. Imposing this constraint during training yielded a test accuracy of $85.33\%$, which is almost comparable to the analog programming scenario. Further reduction in the accuracy gap can be achieved by engineering the grain size of the ferroelectric material \cite{lee2021domains}, resulting in increased number of domains. 

In summary, the intrinsic non-linear synaptic plasticity and ultra-low programming energies of ferroelectric field effect transistors make them an ideal candidate for future brain-inspired computing paradigms that are able to learn unsupervised representations in an online fashion from limited training data.

\vspace{-8mm}

\section*{}

The authors would like to acknowledge GlobalFoundries Dredsen Germany for providing FeFET testing devices. This work was supported as part of the Center for 3D Ferroelectric Microelectronics (3DFeM), an Energy Frontier Research Center funded by the U.S. Department of Energy (DOE), Office of Science, Basic Energy Sciences under Award \#DE-SC0021118.

\section*{Data Availability Statement}
The data that support the findings of this study are available from the corresponding authors upon reasonable request.

%

\end{document}